\documentclass[conference,a4paper]{IEEEtran}
\IEEEoverridecommandlockouts
\usepackage{cite}
\usepackage{amsmath,amssymb,amsfonts}
\usepackage{algorithmic}
\usepackage{graphicx}
\usepackage{textcomp}
\usepackage{xcolor}
\usepackage{epsfig}
\usepackage{bbm}
\usepackage{dsfont}
\usepackage{caption}
\usepackage{subcaption}
\usepackage{verbatim}
\usepackage[absolute]{textpos}
\TPMargin{5pt}

\usepackage{color}

\newcommand{\vbeta}{\boldsymbol{\beta}}

\newcommand{\vq}{\mathbf{q}}

\newcommand{\vx}{\mathbf{x}}
\newcommand{\vy}{\mathbf{y}}

\def\BibTeX{{\rm B\kern-.05em{\sc i\kern-.025em b}\kern-.08em
    T\kern-.1667em\lower.7ex\hbox{E}\kern-.125emX}}
    
\newcommand{\copyrightstatement}{
    \begin{textblock}{0.84}(0.08,0.93)    
         \noindent
         \footnotesize
©2021 IEEE Personal use of this material is permitted. Permission from IEEE must be obtained for all other uses, in any current or future media,
including reprinting/republishing this material for advertising or promotional purposes, creating new collective works, for resale or redistribution to
servers or lists, or reuse of any copyrighted component of this work in other works.   \end{textblock}
}

\begin{document}
\copyrightstatement

\title{Probabilistic Forecast Combination for Anomaly Detection in Building Heat Load Time Series
\thanks{This work is funded by the German federal Ministry for Economic Affairs and Energy. Funding code: 03ET1638.}
}

\author{\IEEEauthorblockN{Mario Beykirch, Tim Janke, Imed Tayeche and Florian Steinke}
\IEEEauthorblockA{\textit{Energy Information Networks and Systems} \\
\textit{TU Darmstadt} \\
Darmstadt, Germany\\
\{mario.beykirch, tim.janke, florian.steinke\}@eins.tu-darmstadt.de, imedtayech@hotmail.fr} 
}

\maketitle
\IEEEpubidadjcol
\begin{abstract}
We consider the problem of automated anomaly detection for building level heat load time series. An anomaly detection model must be applicable to a diverse group of buildings and provide robust results on heat load time series with low signal-to-noise ratios, several seasonalities, and significant exogenous effects.
We propose to employ a probabilistic forecast combination approach based on an ensemble of deterministic forecasts in an anomaly detection scheme that classifies observed values based on their probability under a predictive distribution. 
We show empirically that forecast based anomaly detection provides improved accuracy when employing a forecast combination approach.

\end{abstract}


\section{Introduction}
With further integration of electric and heat energy systems and the increased use of smart meters on building level, the analysis of heat load time series plays an increasing role in local smart energy grids. 
These time series can be utilized, for example, in load forecasts for scheduling problems or to identify and reduce energy waste. 
One important part in energy monitoring is the detection of anomalies, such as unusual consumption, system failures, or measurement errors. 
For large scale monitoring infrastructures, in our case 40 university campus buildings with multiple measured energy forms, a manual anomaly detection is hardly feasible on a daily basis.
This makes an automated anomaly detection necessary.

The literature provides various methodologies for anomaly detection in building energy time series, see \cite{Himeur_ADbuildingLoadReview} for an extensive review.
Several works present building level anomaly detection methods based on probabilistic load forecasts.
The main idea is to use a probabilistic model to issue a predictive distribution over the expected load and classify an observed value as anomalous if this value is unlikely under the predicted distribution.
The authors of \cite{HongYue_RealTimeAD} apply a dynamic regression model to predict the load one hour ahead and select an anomaly threshold from observed percentage errors.
In \cite{ChouTelaga} the load is predicted with an autoregressive neural network and anomaly thresholds are derived from a constant Gaussian error.
\cite{Chen_adGamArch} first predict the day-ahead load with a generalized additive model (GAM) based on exogenous variables and then fit an autoregressive conditional heteroscedasticity (ARCH) model on the forecast residuals to obtain a conditional distribution over the forecast errors.

However, these works do not address the specific challenges associated with anomaly detection in heat load time series on the building level: A considerable number of observed values is zero or close to zero, seasonalities and exogenous variables, such as the temperature, have a significant effect on the load, and by considering single buildings, random effects, such as user behavior, have a large influence on the observed load.
Furthermore, the model must be flexible enough to model a diverse set of buildings.
Finally, the assumption of a standard homoscedastic or heteroscedastic Gaussian error distribution is too simplistic for a reliable anomaly detection.

Our proposed anomaly detection methodology consists of three steps.
First, we create an ensemble of diverse point forecasting models based on different regression techniques and training set sizes.
The forecasts of these models serve as inputs for probabilistic forecast combination techniques to obtain a predictive distribution conditional on the ensemble predictions.
These distributions are then used to classify the observed load values into normal and anomalous based on a selected interval.
This also allows to tune the aggressiveness of the anomaly detection by varying the size on the non-anomaly interval.

The efficacy of our method relies on the quality of the predictive distributions produced by the selected forecasting model.
The classic approach to the model selection problem is to pick a single  model ex-ante based on validation set performance.
This might be computationally expensive, especially when one wants to forecast many time series, and requires large data sets to reliably estimate the out-of-sample accuracy.
The latter point is especially true for building level energy time series as they are usually characterized by a low signal-to-noise ratio and show daily, weekly, and yearly seasonality.
Furthermore, the assumption that a single best model for a time series exists at all is questionable and different models might perform well for different regions of the input space.
This motivates our forecast combination approach.
While the the combination of point forecasts is a well researched topic, probabilistic forecast combination is considered a current frontier in energy forecasting \cite{EnergyForecastingReview}.
To our knowledge this work is the first to propose an anomaly detection methodology based on probabilistic load forecast combination techniques.

We empirically test our approach on the heat load time series of two real university campus buildings.
To quantify the anomaly detection performance, we insert artificial anomalies into the time series.
We find that a carefully designed probabilistic forecast combination model based on generalized additive models for location, scale, and shape (GAMLSS) \cite{GAMLSS_2} improves over simple averaging approaches and individual models not only in terms of the statistical forecast error measures but also leads to significant improvements in anomaly detection performance.

The remainder of the paper is structured as follows.
In the next section we formalize the problem and introduce the necessary notation.
We then present our ensemble of point forecasting models and the considered probabilistic forecast combination approaches.
The setup and results of our empirical study are explained in Section \ref{Sec:EmpiricalEvaluation}.
We conclude in Section \ref{Sec:Conclusion}.

\section{Methodology}
\subsection{Problem definition}
The anomaly detection should answer the question whether an observed load value $y_{t}$ at time step $t$ is an anomaly given the vector of previous observations $\vy_l$ and a vector of explanatory variables $\vx_t$. 



The core idea of forecast based anomaly detection is to classify observed values based on the probability to observe them under a distribution $p$ issued by a probabilistic regression model $f$.
Given the distribution $p$, we can compute the probability that the random variable $Y$ lies in the interval $(a,b]$ as $Pr(a<Y \leq b) = F_Y(b)-F_Y(a)$, where $F_Y$ is the cumulative distribution function (CDF) of the distribution $p$.

Based on this, we define an anomaly detection model that, given a predictive CDF $\hat{F}(y_t)$, classifies $y_t$ as an anomaly if
\begin{equation}
    \hat{F}(y_t) <\tau_{lower} \vee \hat{F}(y_t) >\tau_{upper}
\end{equation}
holds true, with $\tau_{lower}$ and $\tau_{upper}$ being lower and upper threshold levels of a plausibility interval.
To this end, we first need to estimate the CDF $\hat{F}_t$ of  the conditional distribution $p(y_t|\vy_l,\vx_t)$.
If $p$ belongs to a parametric family, $\hat{F}$ is usually available in closed form.
Alternatively, $\hat{F}$ could be approximated by a set of predicted quantiles.

In case of forecast combination, we take a two step approach to estimate $\hat{F}$.
We use a set of $M$ point forecasting models $\{f_m\}_{m=1}^M$ from which we can obtain predictions $\hat{y}_{m,t} = f_m(\vy_l,\vx_t)$. We then use the vector of predictions $\mathbf{\hat{y}}_t=[\hat{y}_{1,t}, ... , \hat{y}_{m,t} ]$ in a second model to obtain $\hat{F}_t=g(\mathbf{\hat{y}}_t)$.

We present the deterministic and probabilistic models in the following.
\subsection{Deterministic forecast ensemble}
We arrange the deterministic forecast ensemble $\mathbf{\hat{y}}_t$ with $M=9$ forecasts. 
These are yielded from three different forecasting methods, each of which is trained with a 60, a 90, and a 365 day training window of size $D_{train}$, consisting of $N_{train}=24 D_{train}$ training samples. 
The effectiveness of combining forecasts of different training window sizes is shown in \cite{Hubicka_caliWindows}, where it is applied to electricity price forecasting. 
We selected the forecasting methods based on their ability to flexibly model load time series of a diverse group of buildings. 

To obtain the forecast time series we re-train the models each day to predict the hourly load of the complete next day, with the training window consisting of the directly preceding $D_{train}$ days, i.e. we perform day-ahead forecasting with a rolling training window.

For each model, we consider one hyper-parameter as tuning parameter. We separately employ a grid search to determine the tuning parameter value that minimizes the mean absolute error (MAE) in a rolling window cross-validation.

We present the three employed deterministic forecast methods in the following. 
\subsubsection{Lasso linear regression (Lasso)}
The first method is a linear regression model with Lasso regularization \cite{tibshirani_ogLasso}. It is commonly applied in the forecasting literature, e.g. to electric load forecasting in \cite{ziel_LassoInLFC}. Lasso regression performs automatic feature selection and parameter shrinkage, which allows us to flexibly model all building time series without ex-ante assumptions about feature relevancy.
We model the load as
\begin{equation}
    \begin{split}
    y_t =& \beta_0 + \sum_{i=1}^7 \left( \beta _i y_{t-24i} \right) + \beta_8 T_t + \beta_9  T_{t-24} + \beta_{10} T^{peak}_{t-24} \\
    &+ \beta_{11} T^{avg} + \beta_{12} y^{peak}_{t-24} + \beta_{13} y^{avg}_{t-24} + \beta_{14} HDH^{avg}_{t-24} \\
    &+ \beta_{15} HDH_{t}^{avg} + \beta_{16} HDH_{t-24}  + \beta_{17} WD_t  \\
    &+\beta_{18} WT_t  + \varepsilon_t,
    \end{split}
\end{equation}
where $\beta_0$ is an intercept coefficient, $\vbeta = [\beta_1,...,\beta_k]$ the regression coefficients, $y_t$ the observed load, $T_t$ the temperature, and $\varepsilon_t$ the model error, each at time step t. We define the heating degree hour as $\mathrm{HDH}_t=\mathrm{max}(18^{\circ}\mathrm{C}-T_t,0)$, which is a linearization of the non-linear temperature effect. Superscripts \textit{avg} and \textit{peak} denote the average and the peak value of the respective day. $WD_t$ and $WT_t$ are binary variables that represent working days and working hours, which we define as from 9:00 to 17:00. 
The model is trained to minimize $\frac{1}{N_{train}}||\boldsymbol{\varepsilon}||_2^2+\lambda_{lasso} ||\vbeta||_1 $.
We consider the lasso parameter $\mathbf{\lambda}_{lasso}$ as the tuning parameter. 
\subsubsection{Gradient boosting regression (GBR)}
The second point forecasting method are gradient boosted regression trees. This method combines an ensemble of weak estimators, in this case decision trees, to form a strong estimator. The tree structure allows data-driven modeling of feature interactions \cite{Friedmann_ogGB}. \cite{Zhang_GB-LFC} apply GBR to electricity load forecasting. 
We consider the feature vector
\begin{equation}
    \begin{split}
    \mathbf{x}^{GBR}_{t} = (& y_{t-24},y_{t-48},y_{t-72}, y_{t-168}, y^{peak}_{t-24},y^{avg}_{t-24}, \\
    & T_t, T_{t-24},T^{peak}_{t-24},HDD_t^{avg},HDD_{t}, \\
    & HOD_t, DOW_t,WOY_t  ),
    \end{split}
\end{equation}
of which $HOD_t$, $DOW_t$, and $WOY_t$ are integer values that represent the hour of the day, the day of the week, and the week of the year, respectively.

We use 300 estimators, a learning rate of $0.1$, and minimize the least squares loss in the training. We allow the maximum tree depth to be between 3 and 6 and consider it the tuning parameter.  

\subsubsection{Generalized additive model (GAM)}
The third method is based on a GAM, which allows flexible modeling of non-linear feature effects with penalized b-spline functions \cite{Hastie_GAMogBook}. GAMs are applied to electric load forecasting for example in \cite{Fan_GAMinLoadFC}. 

The model reads as
\begin{equation}
    \begin{split}
        y_t = & \beta_0 +f_1(y_{t-24}) +f_2(y_{t-168}) +f_3(y_{t-24}^{peak}) \\
              & +f_4(T_{t})+f_5(T^{avg}_{t-24})  + f_6(HOD_t)  + f_7(WOY_t)   \\
              & + \sum_{i=1}^{6}\beta_i DOW_{i,t}  +\varepsilon_t,\\
    \end{split}
\end{equation}
where functions $f_1,...,f_7$ are penalized b-spline functions, of which $f_1,...,f_5$ are defined with 10 b-splines and $f_6$ with 24 b-splines. We only add $f_7$, defined with 5 b-splines, in the 365 days model. $DOW_t$ is expressed as a set of six dummy variables. Functions $f_1$, $f_2$, and $f_3$ are constrained to be monotonically increasing, $f_4$ and $f_5$ to be monotonically decreasing. The model considers the squared error in the training. The different number of splines mitigates over-fitting of feature effects. The smoothing parameter $\lambda_{GAM}$ is set equal for each function and considered the tuning parameter.

The mean forecast is the hourly mean of the ensemble forecast  $\bar{y}_{ens,t}=\frac{1}{M} \sum_{m = 1}^{M}\hat{y}_{m,t}$. 
\subsection{Probabilistic forecast combination}
We compare four different probabilistic forecast combination approaches.

\subsubsection{Ensemble average (EA)}
This model assumes a Gaussian distribution centered at the ensemble mean forecast $\bar{y}_{ens,t}$ and a constant variance equal to the variance of the in-sample residuals of this mean forecast $\bar{\sigma}^2=\frac{1}{N_{train}-1}\sum_{t=1}^{N_{train}} (y_t-\bar{y}_{ens,t})^2$, i.e. the predictive distribution would read $\mathcal{N}(\bar{y}_{ens,t}, \bar{\sigma}^2)$. 
However, since $y_t \ge 0$, such a standard Gaussian distribution would assign non-zero probabilities for physically impossible values below zero.
Therefore, we employ a mixture distribution where the probability mass for negative values is reassigned to a point mass at $0$.
We denote this mixture distribution as $\mathcal{N}^{\mathbf{0}}(\bar{y}_{ens,t}, \bar{\sigma}^2)$.
Note that this is different from a zero-truncated distribution where the probability mass is assigned proportionally to all values $>0$.
See e.g. \cite{Nowotarski_DFC-Combination} for an application to electricity load forecasting.

\subsubsection{Ensemble average and ensemble variance (EA-EV)}
The ensemble variance $s_t^2 = \frac{1}{M-1} \sum_{j=1}^{M}(\bar{y}_{t,ens}-y_j)^2 $ can be interpreted as a proxy for the uncertainty of the mean prediction and can be used to model a distribution with conditional variance.
Hence, the predictive distribution of this model is given by $\mathcal{N}^{\mathbf{0}}(\bar{y}_{ens,t}, s_t^2)$.
However, we expect the predictive distribution of this model to be under-dispersed, i.e. the predicted variance is likely too small on average.

\subsubsection{GAMLSS-based forecast combination (GAMLSS)}
GAMLSS \cite{GAMLSS_2} are univariate distributional regression models where all parameters of the assumed distribution for the response can be modeled as additive functions of the explanatory variables.
The GAMLSS framework also allows to fit censored regression models.
Such models are appropriate when the target variable is naturally censored above and/or below a certain value, see e.g. \cite{Messner2014} for an application to probabilistic wind speed and precipitation forecasting.

We assume that the heat demand follows a t-distribution with a point mass at zero denoted by $y_t \sim t^{\mathbf{0}}(\mu_t, \sigma_t^2, \nu)$.
The t-distribution accommodates the large signal-to-noise ratio of the time series given its ability to model heavy tails when the degrees of freedom parameter $\nu$ is small.
The mean model is a linear combination of the ensemble forecasts
\begin{equation}
    \hat{\mu}_t = \beta_{0} + \sum_{m=1}^{M} \beta_{m} \hat{y}_{m,t}
\end{equation}
and the log-standard deviation is modeled as a nonlinear function of the ensemble standard deviation
\begin{equation}
    log(\hat{\sigma}_t) = \beta_{0} + f(s_t),
\end{equation}
where $s_t$ is the standard deviation of the ensemble and $f$ is a penalized b-spline function with 20 equidistant knots that is constrained to be monotonically increasing.
The degrees of freedom $\nu$ are estimated as a constant.
Parameters are estimated via maximum likelihood using a 0-censored t-likelihood.

\begin{figure*}[h!]
     \centering
     \begin{subfigure}[]{0.49\textwidth}
	\centering
	\includegraphics[width=0.95\linewidth]{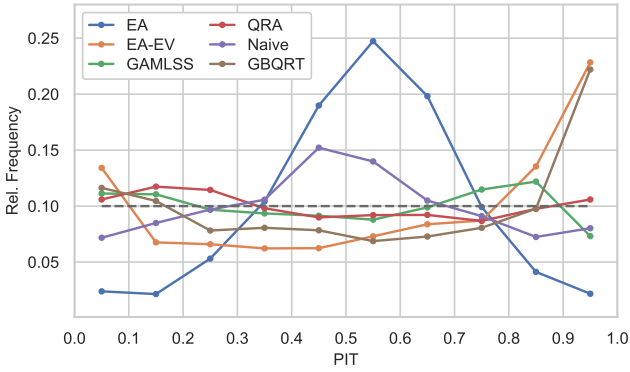}
	\caption{Building A}
	\label{fig:results_PIT_A}
\end{subfigure}
\begin{subfigure}[]{0.49\textwidth}
	\centering
	\includegraphics[width=0.95\linewidth]{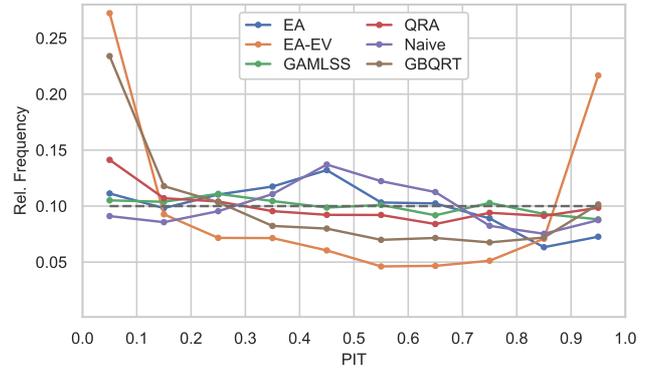}
	\caption{Building B}
	\label{fig:results_PIT_B}
\end{subfigure}
\caption{PIT histograms of each probabilistic forecast in the test period. The points indicate the bin value and are positioned in its center, with each bin having a width of $0.1$. The dashed line represents an optimal calibration.}
\label{fig:results_PIT}
\end{figure*}

\subsubsection{Quantile Regression Averaging (QRA)}
Quantile regression \cite{Koenker_QRAog} offers a flexible approach to probabilistic forecasting as it allows to estimate the $\tau$-quantile $q_{\tau}$ of a CDF without making parametric assumptions about the underlying distribution.
Hence, we can approximate the CDF of a conditional predictive distribution by a vector of $K$ quantiles $\vq=[q_{\tau_1}, ..., q_{\tau_K}]$.

QRA \cite{NowotarskiQRA} is a forecast combination technique that builds on this idea.
The predictive quantile $\hat{q}_{t, \tau}$ for level $\tau$ at time $t$ is given by a linear combination of the ensemble forecasts
\begin{equation}
    \hat{q}_{t, \tau} = \beta_{0,\tau} + \sum_{m=1}^{M} \beta_{m,\tau} \hat{y}_{m,t}.
\end{equation}
Note that one has to estimate a separate model for each quantile, i.e. we estimate 99 models, one for each $\tau \in \{0.01,0.02,...,0.99\}$, to obtain $\hat{\vq}_t$.
We sort the estimated quantiles to ensure monotonicity and clip all negative values.
A predictive CDF $\hat{F}_t$ can then be obtained by interpolation of $\hat{\vq}_t$.
QRA has been applied to to the problem of electrical load forecasting in \cite{Weron_QRAinLoadFC}.

We additionally test two benchmark models that do not rely on forecast combination.
\subsubsection{Naive}
This model is a naive baseline and uses the observed demand of the last day $\hat{y}_{naive,t}=y_{t-24}$ as the mean forecast.
The constant variance is estimated using the observed residuals of this mean forecast, i.e. $\hat{\sigma}_{naive}^2 = \frac{1}{N_{train}-1}\sum_{t=1}^{N_{train}} (y_t-\hat{y}_{naive,t})^2$.
The predictive distribution reads $\mathcal{N}^{\mathbf{0}}(\hat{y}_{naive,t}, \hat{\sigma}_{naive}^2)$.

\subsubsection{Gradient Boosted Quantile Regression Trees (GBQRT)}
As we observed the GBR model with a 365 day training window to be one of the best point forecasting models, we chose this model as a strong benchmark for the forecast combination approaches.
The model uses the same features and hyper-parameters as the point forecasting model but minimizes the quantile loss instead of the squared error.
As for the QRA model, negative values are clipped and the quantiles are sorted in ascending order.

All probabilistic models are re-estimated daily with a rolling window of 365 days.

\subsection{Benchmarking metrics}
We benchmark point forecasts with the  $\mathrm{MAE}=\frac{1}{N_{test}}  \sum_{t=1}^{N_{test}}|y_t-\hat{y}_t^+|$ and the root mean squared error $\mathrm{RMSE}=\sqrt{\frac{1}{N_{test}} \sum_{t=1}^{N_{test}}(y_t-\hat{y}_t^+)^2}$, where $N_{test}$ is the number of samples in the test set and $\hat{y}_t^+=\mathrm{max}(\hat{y}_t,0)$ a point forecast clipped at 0.
To compare the performance of the probabilistic forecasts we use the continuous ranked probability score (CRPS), which is defined as
\begin{equation}\label{eq:crps}
    \mathrm{CRPS}(\hat{F}_t,y_t) =  \int_{-\infty}^{\infty} ( \hat{F}_t(z)- \mathbbm{1} \{y_t \ge z\})^2 \mathrm{d}z. 
\end{equation}
Closed form for solutions for (\ref{eq:crps}) are available for some distributions such as the Gaussian.
Alternatively, the CRPS can be estimated based on samples as
$\mathrm{CRPS}(\hat{F}_t,y_t) = \frac{1}{S}\sum_{i=1}^S |\hat{Y}_{t,i}-y_t| - \frac{1}{2S^2} \sum_{i=1}^S \sum_{j=1}^S |\hat{Y}_{t,i}-\hat{Y}_{t,j}|$
where $\{\hat{Y}_{t,i}\}_{i=1}^S$ is a set of $S$ independent samples from the predictive distribution \cite{Gneiting_Katzfuss}. 
We use $S=1000$ samples for each observation to estimate the CRPS and report the average value over the test period.

We benchmark the anomaly detection by true positive rate $TPR=\frac{TP}{TP+FN}$ and false positive rate $FPR=\frac{FP}{FP+TN}$. TP, FP, FN, and TN are the number of true positive, false positive, false negative, and true negative classifications, respectively.




\begin{figure*}[h!]
     \centering
     \begin{subfigure}[]{0.49\textwidth}
	\centering
	\includegraphics[width=1\linewidth]{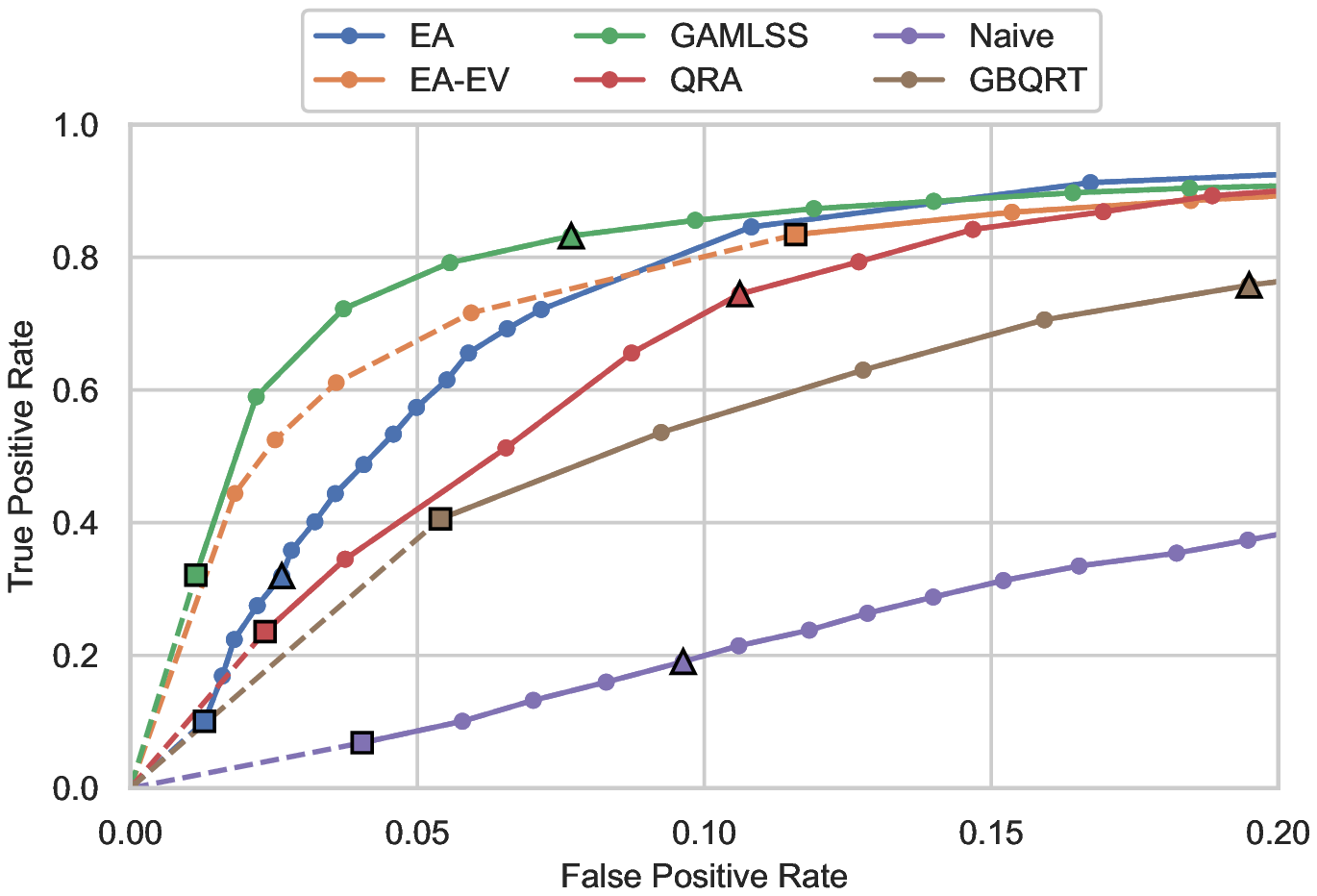}
	\caption{Building A}
	\label{fig:results_ROC_A}
     \end{subfigure}
    \begin{subfigure}[]{0.49\textwidth}
	\centering
	\includegraphics[width=1\linewidth]{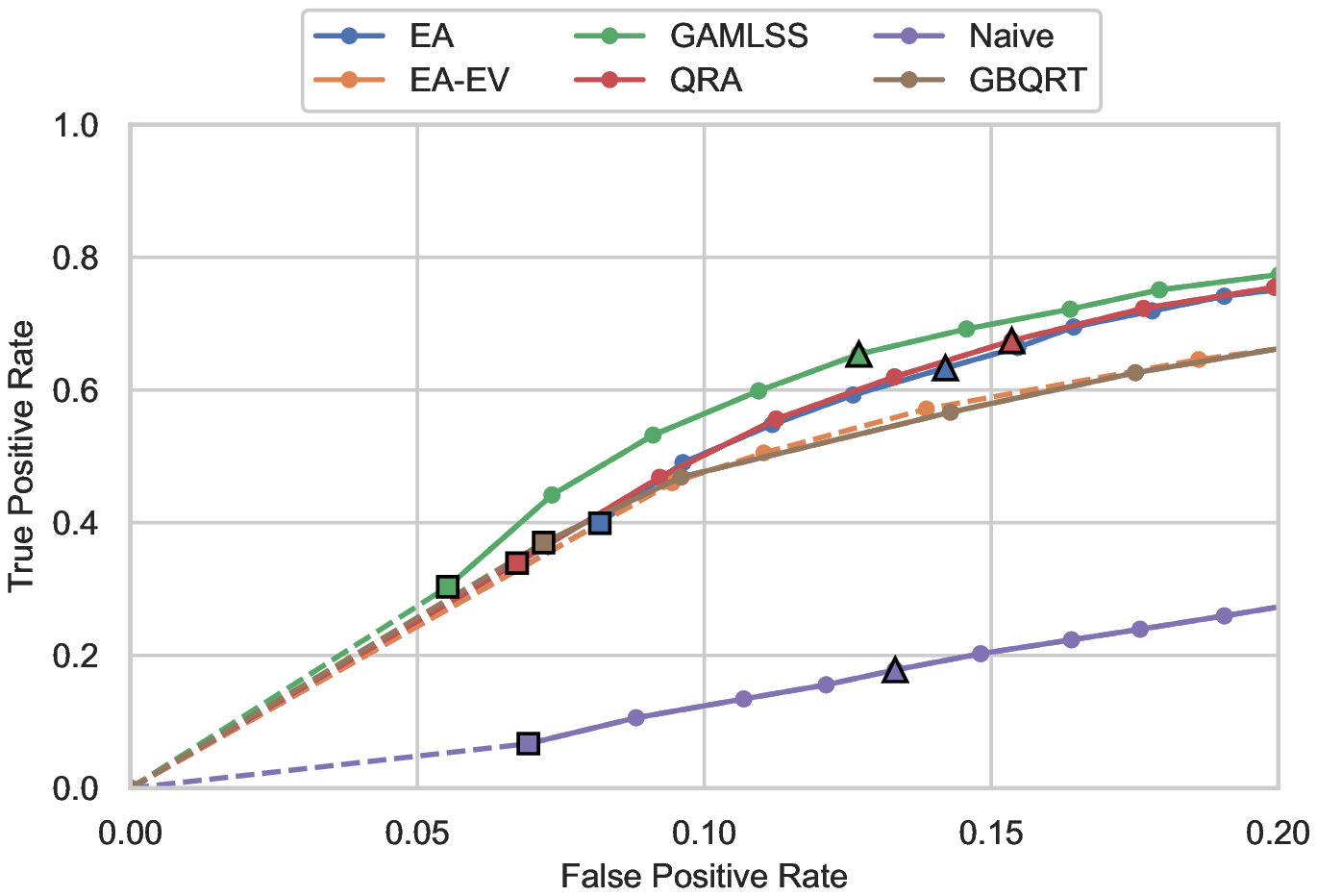}
	\caption{Building B}
	\label{fig:results_ROC_B}
     \end{subfigure}

     \hfill
        \caption{ROC curves for each anomaly detection model with each point representing the result for one quantile level as threshold. Square and triangle mark the thresholds $\tau_{lower}=0.01$ and  $\tau_{lower}=0.05$, respectively. Dashed lines represent quantile levels $\tau_{lower}<0.01$.  The results are averages of 30 runs.}
        \label{fig:results_ROC}
\end{figure*}

\begin{table}[hb!]
\begin{center}
	\caption{MAE and RMSE of each point forecast in the test set. The lowest values in each column is marked in bold. The three best ensemble forecasts, per building and error measure, are marked by a superscript.}
	\begin{tabular}{c c | l l | l l }
		& &  \multicolumn{2}{c|}{Building A }  & \multicolumn{2}{c}{Building B} \\
		\hline
	    Days & Model  						& RMSE 	&  MAE &  RMSE 	&  MAE	\\
		\hline
        60  & Lasso                                 &  55.8 &  35.7  &  17.5$^{(1)}$ &  12.1$^{(1)}$ \\
        60  & GBR                                   &  56.5 &  32.0$^{(3)}$ &   17.6$^{(3)}$ &  12.2$^{(2)}$  \\ 
        60  & GAM                                   &  59.5 &  37.8  &   19.8 &  13.7  \\
        \hline
        90  & Lasso                                 &  52.7$^{(2)}$ &   32.9   &  18.2 &  12.5 \\
        90  & GBR                                   &  60.9 &  31.8$^{(2)}$  &   17.5$^{(1)}$ &  12.2$^{(2)}$ \\
        90  & GAM                                   &  58.3 &  36.3 &  20.2 &  14.1 \\
        \hline
        365 & Lasso                                 &  53.0$^{(3)}$ &  34.5  &  19.2 &  13.3  \\
        365 & GBR                                   & 49.3$^{(1)}$ &  30.4$^{(1)}$ &   18.7 &  13.0 \\
        365 & GAM                                   &  57.3 &  39.9&   21.2 &  14.5 \\
        \hline
        \multicolumn{2}{c|}{Mean forecast}    &  \textbf{48.6} & \textbf{29.7} &   \textbf{16.7} &  \textbf{11.5}  \\
	\end{tabular}
	\label{tab:results_detForcasta}
\end{center}
\end{table}

\section{Empirical evaluation}\label{Sec:EmpiricalEvaluation}
\subsection{Data \& software}
We apply our methodology to the hourly load data of two buildings on the university campus.
The temperature data was obtained from a near-by weather observation station of Deutscher Wetterdienst \cite{DWD}.
We use the data of the year 2017 as the initial training set for the deterministic models.
2018 is the validation set for the deterministic models and the initial training set for the probabilistic models.
2019 is the test set for point forecasts, probabilistic forecasts, and the anomaly detection.
In each year we only consider the heating period, which we define as between September 1st and May 31st.
We don't consider time steps in which an input value or an observed value is missing.
The test set contains 5383 samples for building A and 5340 samples for building B.
In the test period, building A and B have an average load of 394.8kW and 121.9kW, respectively. 

We use the Python packages scikit-learn \cite{scikit-learn}, pyGAM \cite{pyGAM}, and statsmodels \cite{seabold2010statsmodels} as well as the R package gamlss \cite{GAMLSS_2}.

\subsection{Artificial anomaly generation}
To quantify the anomaly detection performance, we replace 5\% of the observed load values in the test period with artificial anomalies.
We create anomalies as deviations of 20\% from the observed load but at least 20\% of the average observed load. 
Otherwise the artificial deviations for very low load values would resemble the usual random fluctuations in the data.
If a deviation results in a negative value, we replace it with a deviation in positive direction.
We run the anomaly detection 30 times.
For each run the artificial anomalies are placed at different random positions and the average performance is reported.
The forecast models are trained, validated, and tested without artificial anomalies.

\begin{table}[b!]
\begin{center}
	\caption{Average test set CRPS values with the lowest per building in bold.}
    \begin{tabular}{ c | cccccc}
        Building & EA & EA-EV & GAMLSS & QRA & Naive & GBQRT\\
    	\hline
        A & 25.51 & 22.00 & \textbf{20.34}  & 21.20   & 62.95   & 21.36\\
        B & 8.55  & 8.88  & \textbf{7.99}   & 8.54    & 21.47   & 9.48 \\
    \end{tabular}
    \label{results_probForcasta}
\end{center}
\end{table}

\subsection{Forecast results}
Table \ref{tab:results_detForcasta} presents the MAE and the RMSE of the point forecasts and the ensemble mean forecast in the test period. 
Notably, different point forecasts of the ensemble perform best for each building and, depending on which error measure we consider, even differ in rank for just one building. The mean forecast, on the other hand, performs best for each building, irrespective of chosen error measure.
This supports our motivation for employing forecast combination and is in line with literature \cite{Nowotarski_DFC-Combination}. 

Table \ref{results_probForcasta} shows the mean test set CRPS score for the probabilistic forecasts.
The GAMLSS model performs best, followed by the QRA model. For building A GBQRT performs better than the simple forecast combination models.

The probability integral transform (PIT) histogram in Figure \ref{fig:results_PIT} shows the relative amount of observed values in equally sized intervals of the predictions' quantile levels and hence provides information about the calibration of the forecasts \cite{Gneiting_Katzfuss}. GAMLSS and QRA models show the best calibration. Calibration in the upper quantile levels suggests that the distributions of the GAMLSS forecast are over-dispersed in the upper half. The other forecasts show worse calibrations, with the EA-EV and GBQRT forecast distributions being under-dispersed and the naive forecast being over-dispersed.


\begin{figure*}[t!]
     \centering
     \begin{subfigure}[]{0.49\textwidth}
         \centering
         \includegraphics[width=1\textwidth]{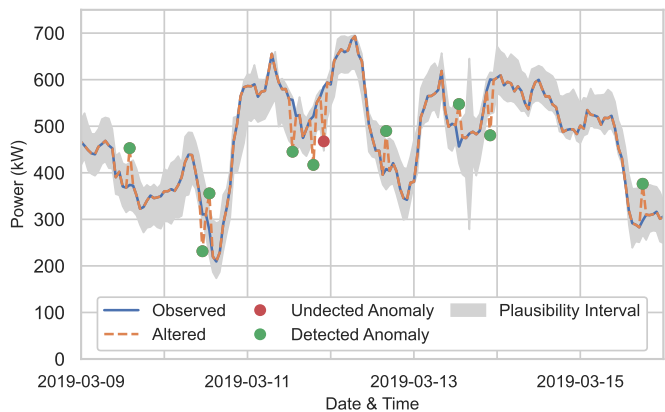}
         \caption{Real load with inserted artificial anomalies}
         \label{fig:results_exampleArtAnoms}
     \end{subfigure}
    \begin{subfigure}[]{0.49\textwidth}
         \centering
         \includegraphics[width=1\textwidth]{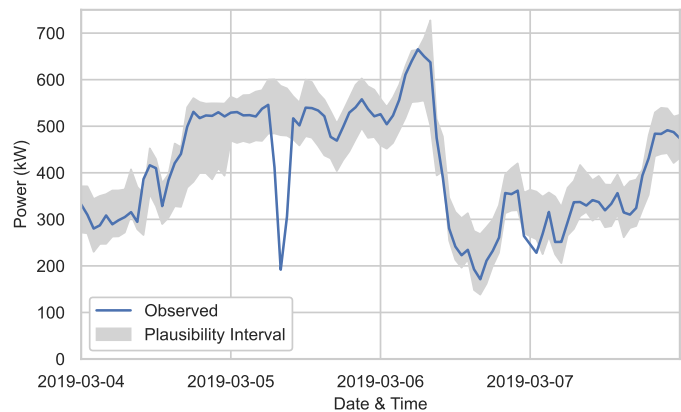}
         \caption{Real load containing an anomaly}
         \label{fig:results_exampleRealAnoms}
     \end{subfigure}
     \hfill
        \caption{Examples of the anomaly detection, when using the GAMLSS model with $\tau_{lower}=0.05$. a) The blue line depicts the observed load from real measurements and the orange line the observed load but altered by inserting artificial anomalies. These are depicted as green dots if detected and red if not. The interval between the anomaly thresholds is depicted in gray. b) The blue line depicts the observed load from real measurements and the gray area depicts the interval between anomaly thresholds.\vspace{-3pt}}
        \label{fig:results_exampleAnoms}
        

\end{figure*}

\subsection{Anomaly detection results}
In the following, when presenting results by threshold quantile levels, we choose thresholds so that $\tau_{upper} = (1-\tau_{lower})$.

The receiver operating characteristic (ROC) curve in Figure \ref{fig:results_ROC_A} shows the TPR vs. the FPR yielding from different anomaly thresholds for each model.
The best anomaly detection maximizes the area under the ROC curve.
The GAMLSS model outperforms the other models and offers interpretable threshold levels.
With good calibration, the quantile level offers an ex-ante interpretable threshold as the FPR should be close to $2\tau_{lower}$.
EA-EV performs better than its CRPS suggests, but due to the inadequate calibration, the quantile levels offer only ex-post interpretability and the threshold levels can't be generalized over different buildings.
All models perform worse on building B, see Figure \ref{fig:results_ROC_B}. The GAMLSS model still performs best.
For both buildings, all models based on forecast combination, except EA-EV for building B, perform better than the benchmark models. 
In summary, a decreased CRPS does not strictly lead to an increased anomaly detection performance. A reason might be that over- and under-dispersion reduce the interpretability of the quantile levels but are not as important when one chooses thresholds ex-post based on the TPR-FPR ratio.
Nevertheless, GAMLSS, with the lowest average CRPS, offers the best anomaly detection for both buildings.

Figure \ref{fig:results_exampleArtAnoms} shows an example of the detection of artificial anomalies. 
The load time series has no recognizable periodicity and the presented anomaly values would be plausible in other time steps. 
Nevertheless, most of the anomalies are detected. 
Figure \ref{fig:results_exampleRealAnoms} shows an example where we know the original data to contain an anomaly. It is correctly detected. 
Considering that a similar steep decline and value is part of a plausible load pattern on the next day, this anomaly can not be trivially identified at the time of measurement. 
This confirms that the proposed methodology offers valuable results in an actual use-case.

\section{Conclusion and Outlook}\label{Sec:Conclusion}
This work presented a novel methodology for anomaly detection in heat load time series based on probabilistic forecast combination techniques.
An empirical study showed that probabilistic forecast combination based anomaly detection outperforms even a strong benchmark.
Of all considered models, the GAMLSS-based forecast combination model yielded the best probabilistic forecast and anomaly detection performance.

Further research could be focused on improving the composition and accuracy of the point forecast ensemble as well as refining the GAMLSS model, e.g. by considering other types of distributions.
The empirical evaluation could be extended to more buildings or other smart meter data sets.
\bibliographystyle{IEEEtran}
\bibliography{conference_101719.bib}
\end{document}